\documentclass[10pt,draftcls,onecolumn]{IEEEtran}
\ifCLASSINFOpdf
\else
   \usepackage[dvips]{graphicx}
\fi
\usepackage{url}
\usepackage{graphicx}
\usepackage{amsmath}
\usepackage{comment}
\usepackage[utf8]{inputenc}
\usepackage[english]{babel}
\usepackage{color}
\usepackage{framed}
\usepackage{amsmath,amssymb,amsfonts}
\usepackage{graphicx}
\usepackage{upgreek}
\usepackage{textcomp}
\usepackage{url}
\usepackage[left=1.62cm,right=1.62cm,top=1.9cm]{geometry}
\usepackage{cite}
\usepackage{bbold}
\usepackage{ifpdf}
\usepackage{times}
\usepackage{layout}
\usepackage{float}
\usepackage{afterpage}
\usepackage{amsmath}
\usepackage{amstext}
\usepackage{amssymb,bm}
\usepackage{latexsym}
\usepackage{color}
\usepackage{flexisym}
\usepackage{kantlipsum}
\usepackage{graphicx}
\usepackage{amsmath}
\usepackage{amsthm}
\usepackage{graphicx}

\usepackage[caption=false,font=footnotesize]{subfig}

\usepackage{booktabs}
\usepackage{multicol}

\usepackage{enumitem}
\usepackage{mathtools}

\usepackage[switch]{lineno}
\usepackage[named]{algo}
\usepackage[noend]{algpseudocode}
\usepackage{algorithm}
\usepackage{amsmath}
\usepackage{algpseudocode}
\usepackage{amsthm}
\usepackage{dsfont}
\usepackage{algorithm}
\usepackage{thmtools}
\usepackage{amssymb}
\usepackage[dvipsnames]{xcolor}

\def\br{\boldsymbol{r}}

\def\bV{\boldsymbol{V}}

\def\bX{\boldsymbol{X}}

\def\bGamma{\boldsymbol{\Gamma}}

\def\bSigma{\boldsymbol{\Sigma}}

\def\bOmega{\boldsymbol{\Omega}}

\def\mba{\mathbf{a}}
\def\mbb{\mathbf{b}}
\def\mbc{\mathbf{c}}

\def\mbr{\mathbf{r}}
\def\mbs{\mathbf{s}}

\def\mbx{\mathbf{x}}
\def\mby{\mathbf{y}}

\def\mbA{\mathbf{A}}
\def\mbB{\mathbf{B}}

\def\mbG{\mathbf{G}}

\def\mbI{\mathbf{I}}

\def\mbP{\mathbf{P}}

\def\mbR{\mathbf{R}}
\def\mbS{\mathbf{S}}

\def\mbV{\mathbf{V}}

\def\mbX{\mathbf{X}}

\theoremstyle{definition}

\algnewcommand\algorithmicinput{\textbf{Input:}}
\algnewcommand\Input{\item[\algorithmicinput]}
\algnewcommand\algorithmicoutput{\textbf{Output:}}
\algnewcommand\Output{\item[\algorithmicoutput]}
\algnewcommand\algorithmicinit{\textbf{Initialize:}}
\algnewcommand\Init{\item[\algorithmicinit]}

\makeatletter
\newcommand*{\rom}[1]{\expandafter\@slowromancap\romannumeral #1@}
\makeatother

\begin{document}
\title{One-Bit Quadratic Compressed Sensing:\\ From Sample Abundance to Linear Feasibility}

\author{%
	\IEEEauthorblockN{Arian Eamaz$^\star$, Farhang Yeganegi$^\star$, Deanna Needell$^{\dagger}$, and Mojtaba Soltanalian$^\star$}
	\\\IEEEauthorblockA{$^{\star}$ECE Department, University of Illinois Chicago, Chicago, USA\\$^{\dagger}$Department of Mathematics, University of California Los Angeles, Los Angeles, USA}
	
	\thanks{The first two authors have contributed equally to this work.}
%	\thanks{* Corresponding author: A. Eamaz (e-mail: aeamaz2@uic.edu).}
}
%\author{Arian Eamaz, Farhang Yeganegi, Deanna Needell, and Mojtaba Soltanalian}

\markboth{%2023 31th European Signal Processing Conference (EUSIPCO)
}
{Shell \MakeLowercase{\textit{et al.}}: Bare Demo of IEEEtran.cls for IEEE Journals}

\maketitle
\thispagestyle{empty}
\pagestyle{empty}
\begin{abstract}
One-bit quantization with time-varying sampling thresholds has recently found significant utilization potential in statistical signal processing applications due to its relatively low power consumption and low implementation cost. In addition to such advantages, an attractive feature of one-bit analog-to-digital converters (ADCs) is their superior sampling rates as compared to their conventional multi-bit counterparts. This characteristic endows one-bit signal processing frameworks with what we refer to as \emph{sample abundance}. On the other hand, many signal recovery and optimization problems are formulated as (possibly non-convex) quadratic programs with linear feasibility constraints in the one-bit sampling regime. We demonstrate, with a particular focus on quadratic compressed sensing, that the sample abundance paradigm allows for the transformation of such quadratic problems to merely a linear feasibility problem by forming a large-scale overdetermined linear system; thus removing the need for costly optimization constraints and objectives. To efficiently tackle the emerging overdetermined linear feasibility problem, we further propose an enhanced randomized Kaczmarz algorithm, called \emph{Block SKM}. Several numerical results are presented to illustrate the effectiveness of the proposed
methodologies.   
\end{abstract}

\IEEEpeerreviewmaketitle

\section{Introduction}
\label{sec:intro}
In the past two decades, sparsity-based processing methods have been attracting a growing interest in statistical signal processing applications\cite{beck2013sparsity}. Quadratic compressed sensing (QCS) is a widely used formulation in sparse signal recovery; examples include when imaging a sparse object using partially and spatially incoherent illumination\cite{shechtman2011sparsity},  or phase retrieval for sparse signals \cite{shechtman2014gespar}. 

To approach the global optimum, the QCS problem was relaxed as a semidefinite programming (SDP) problem, which involves minimizing the rank of a lifted matrix while satisfying both the recovery constraints and the row sparsity constraints on the signal\cite{beck2013sparsity,jaganathan2016phase}. To retrieve the sparse solution, an iterative thresholding algorithm was proposed that leverages a sequence of SDPs. This approach is similar to the recent developments in the field of phase retrieval, where similar semidefinite programming-based ideas have been utilized\cite{jaganathan2013sparse,jaganathan2016phase,candes2013phaselift,candes2014solving}. Unfortunately, these methods have a high complexity, making them difficult to use for the QCS problem. 

To overcome the computational challenges posed by convex optimization techniques, non-convex methods have been introduced as an alternative approach. These methods tackle the phase retrieval problem as a least-square problem and aim to find a local optimum using various optimization techniques\cite{shechtman2014gespar,wang2017solving,bendory2017non}. In\cite{shechtman2014gespar}, they proposed the greedy sparse phase retrieval (GESPAR), a fast local search method, to efficiently recover the signal from measurements of magnitudes in the QCS problem which is more accurate than existing local methods. However, the highly non-convex and non-unique nature of the problem presents a challenge in finding an optimal local solution. To enhance the performance of these local methods, various initialization algorithms have been proposed to improve their outcomes\cite{candes,waldspurger2017phase}.

Sampling the signals of interest at high data rates with high-resolution ADCs would dramatically increase the overall implementation cost and power consumption of the sampling task. In multi-bit sampling scenarios, a very large number of quantization levels is necessary in order to represent the original continuous signal with high accuracy, which in practice, leads to a considerable reduction in sampling rate \cite{eamaz2021modified,boufounos20081}. This attribute of multi-bit sampling has served as a key motivator for the proliferation of underdetermined signal processing tools \cite{candes2013phaselift,candes2015phase,eamaz2022one}. An alternative solution to such challenges is to deploy \emph{one-bit quantization}, which is an extreme sampling scenario, where the signals are merely compared with given threshold levels at the ADC, thus producing sign data ($\pm1$). This enables signal processing equipment to sample at a very high rate, with a considerably lower cost and energy consumption, compared to their conventional counterparts that employ multi-bit ADCs \cite{mezghani2018blind,eamaz2021modified,jacques2013robust,boufounos2015quantization}.

The use of a fixed threshold in one-bit quantization can result in difficulties in accurately estimating the signal amplitude. To address this issue, recent studies have proposed the use of time-varying thresholds, which have been shown to enhance signal recovery performance\cite{eamaz2023covariance,eamaz2022uno,xu2020quantized,laska2011trust}. %These investigations have led to a growing interest in one-bit quantization as a promising approach for high-resolution signal processing. Studies have explored the benefits of time-varying thresholds for various signal characteristics, such as covariance and autocorrelation, and have shown improved estimation results in both stationary and non-stationary scenarios. \cite{eamaz2021modified,qian2017admm,eamaz2022one,eamaz2023covariance,eamaz2022covariance,xi2020gridless}.

In this paper, we consider the deployment of one-bit sampling with time-varying thresholds on QCS, leading to an increased sample size and a \emph{highly overdetermined system} as a result. Our proposed method can recover the desired sparse signal from the \emph{one-bit QCS} by (i) generating abundant one-bit measurements, in order to define a large scale overdetermined system where a finite volume feasible set is created for QCS, and (ii) solving this obtained linear feasibility problem by leveraging one of the efficient solver families of overdetermined systems, namely the \emph{Kaczmarz algorithms}. 

The Kaczmarz method \cite{kaczmarz1937angenaherte} is an iterative projection algorithm for solving linear systems of equations and inequalities. It is usually applied to highly overdetermined systems because of its simplicity.
%Each iteration projects onto the solution space corresponding to one row in the linear system, in a sequential regimen. %The method has been applied to various applications in image reconstruction, digital signal processing, and computer tomography\cite{feichtinger1995kaczmarz,sezan1987applications,eamaz2022one}.
Many variants of this iterative method and their convergence rates have been proposed and studied in recent decades for both consistent and inconsistent systems including the randomized Kaczmarz algorithm, the randomized block Kaczmarz algorithm and most recently, the sampling Kaczmarz-Motzkin (SKM) method \cite{strohmer2009randomized,leventhal2010randomized,needell2014paved,de2017sampling}.

To reconstruct the signal of interest from the one-bit sampled QCS, we employ the novel variant of the Kaczmarz algorithm, \emph{Block} \emph{S}ampling \emph{K}aczmarz-\emph{M}otzkin (Block SKM) whose theoretical guarantees will be discussed.

\underline{\emph{Outline}}:
Section~\ref{QCS} is dedicated to a review of QCS. In Section~\ref{ONEBITQCS}, we will briefly introduce the one-bit sampling via time-varying thresholds and propose the \emph{one-bit polyhedron} for the QCS, which is a large-scale overdetermined system. An accelerated Kaczmarz approach is proposed to find the optimal point in the one-bit QCS polyhedron in Section~\ref{Block SKM}. Also, the convergence rate of proposed algorithm is investigated. Section~\ref{num} is devoted to numerical results of the proposed Kaczmarz algorithm to show its recovery performance in one-bit. Also, we compare the performance of proposed algorithm to that of the well-known high-resolution method, GESPAR for the \emph{phase retrieval} scenario, when the rank of middle matrix is one. Finally, Section~\rom{6} concludes the paper. 

\underline{\emph{Notation:}}
We use bold lowercase letters for vectors and bold uppercase letters for matrices. $\mathbb{C}$ and $\mathbb{R}$ represent the set of complex and real numbers, respectively. $(\cdot)^{\top}$ and $(\cdot)^{\mathrm{H}}$ denote the vector/matrix transpose, and the Hermitian transpose, respectively. $\mbI_{N}\in \mathbb{R}^{N\times N}$ and $\mathbf{0}_{N_{1}\times N_{2}}$ are the identity matrix of size $N$ and all-zero matrix of size $N_{1}\times N_{2}$. $\operatorname{Tr}(.)$ denotes the trace of the matrix argument. The Frobenius norm of a matrix $\mbB$ is defined as $\|\mbB\|_{\mathrm{F}}=\sqrt{\sum^{N_{1}}_{r=1}\sum^{N_{2}}_{s=1}\left|b_{rs}\right|^{2}}$ where $\{b_{rs}\}$ are elements of $\mbB$. The $\ell^{0}$-norm of a vector counts the number of its non-zero elements. The Hadamard (element-wise) product of two matrices $\mbB_{1}$ and $\mbB_{2}$ is denoted as $\mbB_{1}\odot \mbB_{2}$. The vectorized form of a matrix $\mbB$ is written as $\operatorname{vec}(\mbB)$. $\mathbf{1}_{s}$ is the $s$-dimensional all-one vector. Given a scalar $x$, we define $(x)^{+}$ as $\max\left\{x,0\right\}$.  The function $\operatorname{sgn}(\cdot)$ yields the sign of its argument. The floor operation is denoted by $\lfloor\rfloor$.
\vspace{-10pt}
\section{Quadratic Compressed Sensing}
\label{QCS}
In QCS, a sparse high-dimensional signal is to be recovered from a quadratic cost function \cite{beck2013sparsity,shechtman2014gespar}:
\begin{equation}
\label{NEG1}
\begin{aligned}
\min_{\mathbf{x}}\quad & \left\|\mathbf{x}\right\|_{0}\\
\text{s.t.} \quad & y_{j}=\mathbf{x}^{\mathrm{H}}\mbA_{j}\mathbf{x},\quad j\in\mathcal{J}=\left\{1,\cdots,m\right\},
\end{aligned}
\end{equation}
where $\mathbf{x}\in \mathbb{C}^{n}$ is the signal to be recovered, $\left\{y_{j}\right\}$ are the measurements, $\left\{\mbA_{j}\right\}\in \mathbb{R}^{n\times n}$ are the associated sensing matrices, and $m$ is the number of measurements. The convex relaxation of (\ref{NEG1}) is obtained by the matrix lifting procedure, given by
\begin{equation}
\label{neg2} 
\begin{aligned}
y_{j} =\mathbf{x}^{\mathrm{H}} \mbA_{j} \mathbf{x}=\operatorname{Tr}\left(\mbA_{j}\mathbf{x x}^{\mathrm{H}}\right)=\operatorname{Tr}\left(\mbA_{j} \mbX\right),
\end{aligned}
\end{equation}
where $\mbX=\mathbf{x x}^{\mathrm{H}}$, and $\operatorname{Tr}\left(\mbA_{j}\mbX\right)=\operatorname{vec}\left(\mbA^{\top}_{j}\right)^{\top}\operatorname{vec}\left(\mbX\right)$. The sparsity constraint on $\mathbf{x}$ can be dealt with by enforcing the row-sparsity of $\mbX$. If $\mathbf{x}$ has $k$ non-zero elements, then $\mbX$ has $k$ rows containing non-zero elements, and each of these rows is also $k$-sparse. The row-sparsity of $\mbX$ may be promoted by adding a quadratic constraint on $\mbX$, i.e., $\sum_{r}\left(\sum_{s}\left|\mbX_{rs}\right|^{2}\right)^{\frac{1}{2}}<\eta$, where $\eta$ is a positive number\cite{shechtman2011sparsity}. Based on (\ref{neg2}), the QCS problem can be reformulated as:
\begin{equation}
\label{neg3}
\begin{aligned}
\text{find}\quad &\bX\\
\text{s.t.} \quad &\operatorname{Tr}\left(\mbA_{j} \mbX\right)=y_{j}, \\
&\sum_{r}\left(\sum_{s}\left|\mbX_{rs}\right|^{2}\right)^{\frac{1}{2}}<\eta,\\& \operatorname{rank}\left(\mbX\right)=1,~
\mbX \succeq 0.
\end{aligned}
\end{equation}
To have a convex program similar to \cite{shechtman2011sparsity}, the problem (\ref{neg3}) may be relaxed as,
\begin{equation}
\label{neg4}
\begin{aligned}
\min_{\mbX}\quad &\operatorname{Tr}(\mbX)\\
\text{s.t.} \quad &\operatorname{Tr}\left(\mbA_{j} \mbX\right)=y_{j}, \\&\sum_{r}\left(\sum_{s}\left|\mbX_{rs}\right|^{2}\right)^{\frac{1}{2}}<\eta,~\mbX \succeq 0.
\end{aligned}
\end{equation}
The above problem is a semi-definite program (SDP). Similar SDP-based ideas were recently utilized in the context of phase retrieval \cite{candes2013phaselift,candes2015phase}. However, the SDP has a high computational complexity; in particular, the semi-definiteness and row-sparsity constraint in the above problem render it computationally demanding \cite{luo2010semidefinite,naghsh2013doppler,shechtman2014gespar}. 

An interesting alternative to enforcing the \emph{feasible set}  in problem (\ref{neg4}), denoted as $\mathcal{F}_{\mbX}$, emerges when one increases the number of samples $m$, and solves the overdetermined linear system of equations with $m\gg n$. In this sample abundance regimen, the linear constraint $\operatorname{Tr}\left(\mbA_{j}\mbX\right)=y_{j}$ may actually yield the optimum inside $\mathcal{F}_{\mbX}$. As a result of increasing the number of samples, it is possible that the intersection of these hyperplanes will achieve the optimal point without the need to consider other costly constraints. However, this idea may face practical limitations in the case of multi-bit quantization systems since ADCs capable of ultra-high rate sampling are difficult and expensive to produce. Moreover, one cannot necessarily expect these constraints to intersect with $\mathcal{F}_{\mbX}$ in such a way to form a finite-volume space before the optimum is obtained \cite{eamaz2022one,candes2013phaselift}.

In the next section, by deploying the idea of one-bit sampling with time-varying thresholds, linear equality constraints are superseded by a massive array of linear inequalities---thus forming a polyhedron that asymptotically coincides with $\mathcal{F}_{\mbX}$.
\vspace{-9pt}
\section{One-Bit QCS}
\label{ONEBITQCS}
In this section, we will briefly introduce the one-bit sampling with time-varying scheme and the signal reconstruction problem in the one-bit quantization scheme. We will demonstrate that the utilization of time-varying thresholds in one-bit sampling results in a highly over-determined system, represented as a polyhedron. Subsequently, by exploiting the \emph{ample of samples} in the one-bit sampling approach, the one-bit sampled QCS problem will be formulated as a \emph{linear feasibility} problem.
\subsection{One-Bit Sampling with Time-Varying Thresholds}
Consider a bandlimited signal $y\in L^{2}$, which is to be represented by its samples via the standard sampling formula\cite{bhandari2020unlimited},
\begin{equation}
\label{eq:1}
0<\mathrm{T} \leqslant \frac{\pi}{\Omega}, \quad y(t)=\sum_{k=-\infty}^{k=+\infty} y(k \mathrm{T}) \operatorname{sinc}\left(\frac{t}{\mathrm{T}}-k\right),
\end{equation}
where $1/\mathrm{T}$ is the sampling rate and $\operatorname{sinc}(t)=\frac{\sin (\pi t)}{(\pi t)}$ is an \emph{ideal} low-pass filter. Suppose $y_{k}=y(k\mathrm{T})$ denotes the uniform samples of $y(t)$ with the sampling rate $1/\mathrm{T}$. Let $r_{k}$ denote the quantized version of $y[k]$ with the formulation $r_{k} = Q(y_{k})$, where $Q$ denotes the quantization effect. In one-bit quantization, compared to zero or constant thresholds, time-varying sampling thresholds yield a better reconstruction performance \cite{ameri2018one,eamaz2022covariance}. These thresholds may be chosen from any distribution. In this work, to be consistent with state-of-the-art \cite{ameri2018one,khobahi2018signal,eamaz2022uno}, we consider a Gaussian non-zero time-varying threshold vector $\boldsymbol{\uptau}=\left[\tau_{k}\right]$ that follows the distribution $\boldsymbol{\uptau} \sim \mathcal{N}\left(\mathbf{d}=\mathbf{1}d,\bSigma\right)$. In the case of one-bit quantization with such time-varying sampling thresholds, the quantizer is simply written as
$r_{k} = \operatorname{sgn}\left(y_{k}-\tau_{k}\right)$. Let $\mathbf{y}=[y_{k}]$ and $\mbr=[r_{k}]$. Then, the signal feasibility based on the one-bit measurements takes the form
\begin{equation}
\label{eq:5}
\mbr \odot \left(\mathbf{y}-\uptau\right) \geq \mathbf{0},
\end{equation}
or equivalently
\begin{equation}
\label{eq:6}
\begin{aligned}
\bOmega \mathbf{y} &\succeq \mbr \odot \uptau,
\end{aligned}
\end{equation}
where $\bOmega \triangleq \operatorname{diag}\left\{\mbr\right\}$. Suppose $\mathbf{y},\uptau \in \mathbb{R}^{m}$, and that $\uptau^{(\ell)}$ denotes the time-varying sampling threshold in $\ell$-th experiment where  $\ell\in\mathcal{L}=\{1,\cdots,m_{1}\}$. According to (\ref{eq:6}), for the $\ell$-th experiment we have
\begin{equation}
\label{eq:7}
\begin{aligned}
\bOmega^{(\ell)} \mathbf{y} &\succeq \br^{(\ell)} \odot \uptau^{(\ell)}, \quad \ell \in \mathcal{L},
\end{aligned}
\end{equation}
where $\bOmega^{(\ell)}=\operatorname{diag}\left\{\mbr^{(\ell)}\right\}$. In (\ref{eq:7}), we have $m_{1}$ linear system of inequalities which can be put together and expressed as
\begin{equation}
\label{eq:8}
\Tilde{\bOmega} \mathbf{y} \succeq \operatorname{vec}\left(\mbR\right)\odot \operatorname{vec}\left(\bGamma\right),
\end{equation}
where $\mbR$ and $\bGamma$ are matrices, with $\left\{\mbr^{(\ell)}\right\}$ and $\left\{\uptau^{(\ell)}\right\}$ representing their columns, respectively, and $\Tilde{\bOmega}$ is given by
\begin{equation}
\label{eq:9}
\Tilde{\bOmega}=\left[\begin{array}{c|c|c}
\bOmega^{(1)} &\cdots &\bOmega^{(m)}
\end{array}\right]^{\top}, \quad \Tilde{\bOmega}\in \mathbb{R}^{m_{1} m\times m}.
\end{equation}
Utilizing the one-bit quantization technique with multiple time-varying sampling threshold sequences allows for an increase in the number of samples with little extra cost and serves as a gateway to the realm of few-bit sampling. This can be especially beneficial in applications where measurement limitations exist.
 
\vspace{-8pt}
\subsection{One-Bit QCS as Linear Feasibility Problem}
Hereafter, we will focus on (\ref{eq:8}) as an overdetermined linear system of inequalities that is associated with the one-bit sampling scheme.
If we apply one-bit sampling to the QCS \eqref{NEG1}, referred to as one-bit QCS,
\begin{equation}
\label{St_5}
r^{(\ell)}_{j}=\begin{cases} +1 &\operatorname{Tr}\left(\mbA_{j} \mbX\right)>\uptau^{(\ell)}_{j}, \\ -1&\operatorname{Tr}\left(\mbA_{j} \mbX\right)<\uptau^{(\ell)}_{j}.\end{cases}
\end{equation}
As a result, by using the linear property of trace function $\operatorname{Tr}\left(\mbA_{j}\mbX\right)=\operatorname{vec}\left(\mbA^{\top}_{j}\right)^{\top}\operatorname{vec}\left(\mbX\right)$, the \emph{one-bit QCS} polyhedron can be written as
\begin{equation}
\label{eq:800n}
\begin{aligned}
\mathcal{P} = \left\{\mbX \mid r^{(\ell)}_{j}\operatorname{vec}\left(\mbA^{\top}_{j}\right)^{\top}\operatorname{vec}\left(\mbX\right)\geq r^{(\ell)}_{j}\uptau^{(\ell)}_{j},~\ell\in\mathcal{L},~ j\in\mathcal{J}\right\},
\end{aligned}
\end{equation}
which is vectorized based on $\mathbf{y}=\mbV\operatorname{vec}\left(\mbX\right)$, where $\mbV$ is a matrix with $\left\{\operatorname{vec}\left(\mbA^{\top}_{j}\right)\right\}$ as its rows.
The inequality (\ref{eq:800n}) may be recast in the standard polyhedron form as
\begin{equation}
\label{eq:8n}
\begin{aligned}
\mathcal{P} = \left\{\mbX \mid \mbP \operatorname{vec}\left(\mbX\right)\succeq \operatorname{vec}\left(\mbR\right)\odot \operatorname{vec}\left(\bGamma\right)\right\},
\end{aligned}
\end{equation}
where $\mbP=\Tilde{\bOmega} \bV$. By leveraging the sample abundance in the one-bit sampling, the space constrained by  (\ref{eq:8n}), \emph{shrinks} to become contained inside the \emph{feasible region}. However, this shrinking space always contains the globally optimal solution, with a volume that is decreasing with an increasing number of one-bit samples.
We will discuss our approach to find the desired matrix $\mbX^{\star}$ below. 
\vspace{-6pt}
\section{Proposed Algorithm}
\label{Block SKM}
To recover the desired signal within the one-bit QCS polyhedron, we use an accelerated variant of randomized Kaczmarz algorithm (RKA). Many variants of this iterative method and their convergence rates have been proposed and studied in recent decades for both consistent and inconsistent systems, including the original randomized Kaczmarz algorithm, the randomized block Kaczmarz algorithm and most recently, the sampling Kaczmarz-Motzkin (SKM) method \cite{strohmer2009randomized,leventhal2010randomized,de2017sampling}. The block-structured nature of the one-bit QCS matrix has motivated the development of the SKM method, designed specifically to handle block-structured linear feasibility problems with efficiency. Further, the proposed algorithm will be backed by theoretical guarantees.
\vspace{-8pt}
\subsection{SKM Method}
The SKM is a \emph{subconjugate gradient method} to solve overdetermined linear systems, i.e., $\mathbf{B}\mathbf{x}  \preceq\mathbf{b}$, where $\mathbf{B}$ is a $m_{1}m\times n$ matrix. The conjugate-gradient methods immediately turn such an inequality to an equality of the following form:
\begin{equation}
\label{eq:10}
\left(\mathbf{B}\mathbf{x}-\mathbf{b}\right)^{+}=0,
\end{equation}
and then approach the solution by the same process as used for systems of equations. 
Given a sample index set $\mathcal{J}$, without loss of generality, rewrite (\ref{eq:10}) as the polyhedron 
\begin{equation}
\label{eq:11}
\begin{aligned}
\begin{cases}
\mathbf{c}_{j} \mathbf{x} \leq b_{j} & \left(j \in \mathcal{I}_{\leq}\right), \\ \mathbf{c}_{j} \mathbf{x}=b_{j} & \left(j \in \mathcal{I}_{=}\right),\end{cases}
\end{aligned}
\end{equation}
where the disjoint index sets $\mathcal{I}_{\leq}$ and $\mathcal{I}_{=}$ partition $\mathcal{J}$ and $\{\mathbf{c}_{j}\}$ are the rows of $\mathbf{B}$. The projection coefficient $\beta_{i}$ of the SKM at $i$ iteration is 
\cite{leventhal2010randomized,briskman2015block,dai2013randomized}
\begin{equation}
\label{eq:12}
\beta_{i}= \begin{cases}
\left(\mathbf{c}_{j} \mathbf{x}_{i}-b_{j}\right)^{+} & \left(j \in \mathcal{I}_{\leq}\right), \\ \mathbf{c}_{j} \mathbf{x}_{i}-b_{j} & \left(j \in \mathcal{I}_{=}\right).
\end{cases}
\end{equation}
The central contribution of SKM lies in its innovative way of projection plane selection. The hyperplane selection is done as follows. At iteration $i$ the SKM algorithm selects a collection of $\gamma$ (denoted by the set $\mathcal{T}_{i}$), uniformly at random out of $m_{1}m$ rows of the constraint matrix $\mbB$. Then, out of these $\gamma$ rows, the row with maximum positive residual is selected.  Finally, the solution is updated as \cite{de2017sampling,sarowar2020sampling}: $\mathbf{x}_{i+1}=\mathbf{x}_{i}-\lambda_{i}\frac{\beta_{i}}{\|\mbc_{_{j^{\star}_{i}}}\|^{2}_{2}} \mbc^{\mathrm{H}}_{_{j^{\star}_{i}}}$, where the index $j^{\star}_{i}$ is chosen as the \emph{Motzkin sampling}, i.e., $j^{\star}_{i}=\operatorname{argmax}~\left\{ \left(\mbc_{j} \mathbf{x}_{i}-b_{j}\right)^{+}\right\},~j\in\mathcal{T}_{i}$ at iteration $i$, and $\lambda_{i}$ is a relaxation parameter which for consistent systems must satisfy,
$0\leq \lim_{i\to\infty} \inf \lambda_{i}\leq \lim_{i\to\infty} \sup \lambda_{i}<2$,
to ensure convergence \cite{strohmer2009randomized}. 
The convergence bound for SKM is given by
\begin{equation}
\label{eq:150}
\mathbb{E}\left\{\left\|\mbx_{i}-\mbx_{\star}\right\|^{2}_{2}\right\} \leq \left(1-\frac{2\lambda_{i}-\lambda^{2}_{i}}{\kappa^{2}\left(\mbB\right)}\right)^{i}~ \left\|\mbx_{0}-\mbx_{\star}\right\|^{2}_{2},
\end{equation}
with $\kappa\left(\mbB\right)=\|\mbB\|_{\mathrm{F}}\|\mbB^{\dagger}\|_{2}$ denoting the scaled condition number, and $\mbx_{\star}$ is the optimal solution.
%In the case where the constraint matrix is normalized, i.e. $\|\mbc_{j}\|^{2}_{2}=1$, $s_{i}$ is the number of satisfied constraints after iteration $i$, and $V_{i}=\max\left\{m-s_{i},m-\gamma+1\right\}$, for the $i$-th iteration we have \cite{de2017sampling},
%\begin{equation}
%\label{eq:1500}
%\mathbb{E}\left\{\left\|\mbx_{i}-\mbx_{\star}\right\|^{2}_{2}\right\} \leq \left(1-\frac{\sigma^{2}_{min}\left(2\lambda_{i}-\lambda^{2}_{i}\right)}{V_{i}}\right)^{i}~ \left\|\mbx_{0}-\mbx_{\star}\right\|^{2}_{2}.
%\end{equation}
%This recovery error bound is tighter than (\ref{eq:150}).
%\vspace{-10pt}
\subsection{Block SKM Algorithm}
The matrix $\mbP$ in (\ref{eq:8n}) has a block structure with the following formulation:
\begin{equation}
\label{eq:90}
\mbP=\left[\begin{array}{c|c|c}
\mbV^{\top}\bOmega^{(1)} &\cdots &\mbV^{\top}\bOmega^{(m)}
\end{array}\right]^{\top}, \quad \mbP\in\mathbb{R}^{m_{1}m\times n}.
\end{equation}
Therefore, it is useful to investigate the accelerated block-based RKA methods to find the desired signal in (\ref{eq:8n}) for further computational efficiency enhancement. Our proposed algorithm, the \emph{Block SKM}, is described as follows.
%\begin{Algorithm}[Block SKM]
%\label{Block_SKM}
Suppose we have a linear feasibility problem $\mbB\mathbf{x}\preceq \mathbf{b}$ where $\mbB=\left[\begin{array}{c|c|c}\mbB^{\top}_{1} &\cdots&\mbB^{\top}_{m_{1}}\end{array}\right]^{\top},$ $ \mbB\in\mathbb{R}^{m_{1}m\times n}$, and $\mathbf{b}=\left[\begin{array}{c|c|c}\mathbf{b}^{\top}_{1} &\cdots&\mathbf{b}^{\top}_{m_{1}}\end{array}\right]^{\top}$. The proposed algorithm for sparse signal recovery, i.e., the Block SKM, may be summarized as follows:
\begin{enumerate}
    \item Choose a block $\mbB_{j}$ uniformly at random with the probability $\operatorname{Pr}\{j=k\}=\frac{\left\|\mbB_{k}\right\|^{2}_{\mathrm{F}}}{\|\mbB\|_{\mathrm{F}}^{2}}$.
    \item Compute $\mathbf{e}=\mbB_{j}\mathbf{x}-\mathbf{b}_{j}$.
    \item Let $\mathbf{e}^{\prime}$ denote the sorted version of $\mathbf{e}$ from $e_{\text{max}}$ (the maximum element of $\mathbf{e}$) to $e_{\text{min}}$ (the minimum element of $\mathbf{e}$). This step is inspired by the idea of the Motzkin sampling, presented in \cite{de2017sampling}, to have an accelerated convergence.
    \item Select the first $k^{\prime}<n$ element of $\mathbf{e}^{\prime}$ and construct the sub-problem $\mbB_{j}^{\prime}\mathbf{x}	\preceq\mathbf{b}_{j}^{\prime}$, where $\mbB_{j}^{\prime}\in\mathbb{R}^{k^{\prime}\times n}$ and $\mathbf{b}_{j}^{\prime}\in\mathbb{R}^{k^{\prime}\times 1}$. The reason behind choosing $k^{\prime}<n$ is due to the computation of $\left(\mbB_{j}^{\prime}\mbB_{j}^{\prime\top}\right)^{-1}$ in the next step (Step~$5$). For $k^{\prime}>n$, the matrix $\mbB_{j}^{\prime}\mbB_{j}^{\prime\top}$ is rank-deficient and its inverse is not available.
    \item Compute the Moore-Penrose of $\mbB_{j}^{\prime}$, i.e.,  $\mbB_{j}^{\prime\dagger}=\mbB_{j}^{\prime\top} \left(\mbB_{j}^{\prime}\mbB_{j}^{\prime\top}\right)^{-1}$.
    \item Update the solution $\mathbf{x}_{i+1}=\mathbf{x}_{i}-\lambda_{i}\mbB_{j}^{\prime\dagger}\left(\mbB_{j}^{\prime}\mathbf{x}-\mathbf{b}_{j}^{\prime}\right)^{+}$. This update process is inspired from the randomized block Kaczmarz method \cite{elfving1980block,needell2014paved} which takes advantage of the efficient matrix-vector multiplication, thus giving the method a significant reduction in computational cost \cite{briskman2015block}.
\end{enumerate}
Particularly, in the case of the one-bit QCS polyhedron, $\mbB=-\mbP$, $\mathbf{x}=\operatorname{vec}\left(\mbX\right)$, and $\mbb=-\operatorname{vec}\left(\mbR\right)\odot \operatorname{vec}\left(\bGamma\right)$.
%\end{Algorithm}
\vspace{-10pt}
\subsection{Convergence Analysis}
%\vspace{-12pt}
It is worth pointing out that the Block SKM algorithm can be considered to be a special case of the more general \emph{sketch-and-project} method, defined as\cite{derezinski2022sharp}:
\begin{equation}
\label{neg_2}
\mbx_{i+1}= \underset{\mbx}{\textrm{argmin}}~\left\|\mbx-\mbx_{i}\right\|^{2}_{2}\quad \textrm{subject to}\quad \mbS^{\top}\mbB\mbx\preceq \mbS^{\top}\mbb,
\end{equation}
where $\mbS\in\mathbb{R}^{m_{1}m\times k^{\prime}}$ is the sketch matrix choosing a block uniformly at random from the main matrix as mentioned in step $1$. The second step of the proposed algorithm follows the Motzkin sampling
where the index $j^{\star}_{i}$ is chosen in $i$-th iteration as follows:
\begin{equation}
\label{St_1}   
j^{\star}_{i}=\underset{j}{\textrm{argmax}}\left\{ \left(\left(\mbS^{\top}\mbB\right)_{j}\mbx_{i}- \left(\mbS^{\top}\mbb\right)_{j}\right)^{+}\right\},
\end{equation}
with $(\cdot)_{i}$ denoting the $i$th row of the matrix argument.

In the Block SKM algorithm, the sketch matrix is given by
\begin{equation}
\label{St_2}
\mbS=\left[\begin{array}{c|c|c}
\mathbf{0}_{k^{\prime}\times p} &\mbI_{k^{\prime}} &\mathbf{0}_{k^{\prime}\times (m_{1}m-k^{\prime}-p)}
\end{array}\right]^{\top}, ~ \mbS\in\mathbb{R}^{m_{1}m\times k^{\prime}},
\end{equation}
where $k^{\prime}$ is the block size and $p=k^{\prime}\alpha,~\alpha\in\left\{1,\cdots,\left\lfloor\frac{m_{1}m}{k\prime}\right\rfloor\right\}$. Note that the literature does not offer any theoretical guarantees for the convergence of the Block SKM with the above sketch matrix~\cite{rebrova2021block}. To derive our theoretical guarantees for the algorithm used to solve the one-bit QCS, we change the sketch matrix to the \emph{Gaussian} sketch matrix as follows:
\begin{equation}
\label{St_3}
\mbS=\left[\begin{array}{c|c|c}
\mathbf{0}_{k^{\prime}\times p} &\mbG &\mathbf{0}_{k^{\prime}\times (m_{1}m-k^{\prime}-p)}
\end{array}\right]^{\top}, ~ \mbS\in\mathbb{R}^{m_{1}m\times k^{\prime}},
\end{equation}
where $\mbG$ is a $k^{\prime}\times k^{\prime}$ Gaussian matrix, whose entries are i.i.d. following the distribution $\mathcal{N}\left(0,1\right)$. In this framework, we are able to provide some theoretical guarantees by taking advantage of the favorable properties of Gaussian random variables.

Assume that $\mathcal{S}$ denotes a non-empty solution set of the polyhedron (\ref{eq:8n}). Owing to the fact that $\mathbb{E}\left\{\left\|\mbx_{i+1}-\mathcal{S}\right\|_{2}^{2}\right\}\leq\mathbb{E}\left\{\left\|\mbx_{i+1}-\mbx_{\star}\right\|_{2}^{2}\right\}$ \cite{leventhal2010randomized}, then we proceed to prove the convergence rate by employing $\mathbb{E}\left\{\left\|\mbx_{i+1}-\mbx_{\star}\right\|_{2}^{2}\right\}$.
Using the fact that $\mbx_{i+1}-\mbx_{\star}$ is orthogonal to $(\mbS^{T}\mbB)_{j^{\star}_{i}}$ \cite{rebrova2021block}, where $j^{\star}_{i}$ is the index chosen based on the Motzkin sampling for the $i$-th iteration, we have the following Pythagorean relation \cite{rebrova2021block,derezinski2022sharp}:
\begin{equation}
\label{St_10}
\begin{aligned}
\left\|\mbx_{i+1}-\mbx_{\star}\right\|_2^2=\left\|\mbx_i-\mbx_{\star}\right\|_2^2-\frac{\left\|\left(\left(\mbS^{\top}\mbB\right)_{j^{\star}_{i}}\mbx_{i}-\left(\mbS^{\top} \mbb\right)_{j^{\star}_{i}}\right)^{+}\right\|_2^2}{\left\|\left(\mbS^{\top} \mbB\right)_i\right\|_2^2}.
\end{aligned}
\end{equation}
In the linear inequality system, the Kaczmarz algorithms only updates the solution when $\mbS^{\top}\mbB\mbx_{i}\succeq \mbS^{\top}\mbb$ at $i$-th iteration. Therefore, one can readily rewrite \eqref{St_10} at iteration $i$ where the condition $\mbS^{\top}\mbB\mbx_{i}\succeq\mbS^{\top}\mbb$ is met:
\begin{equation}
\label{St_100}
\begin{aligned}
\left\|\mbx_{i+1}-\mbx_{\star}\right\|_2^2=\left\|\mbx_i-\mbx_{\star}\right\|_2^2-\frac{\left\|\mbS^{\top}\mbB\mbx_i-\mbS^{\top} \mbb \right\|_{\infty}^2}{\left\|\left(\mbS^{\top}\mbB\right)_{j^{\star}_{i}}\right\|_2^2}.
\end{aligned}
\end{equation}
By taking the expectation over the error, we have
\begin{equation}
\label{St_11}
\mathbb{E}_{\mbS}\left\{\left\|\mbx_{i+1}-\mbx_{\star}\right\|_2^2\right\} = \left\|\mbx_i-\mbx_{\star}\right\|_2^2-\mathbb{E}_{\mbS} \left\{\frac{\left\|\mbS^{\top}\mbB\mbx_i-\mbS^{\top} \mbb \right\|_{\infty}^2}{\left\|\left(\mbS^{\top}\mbB\right)_{j^{\star}_{i}}\right\|_2^2}\right\}.
\end{equation}
In addition, we have that
\begin{equation}
\label{St_12}
\begin{aligned}
\mathbb{E}_{\mbS}\left\{\left\|\left(\mbS^{\top} \mbB\right)_{j^{\star}_{i}}\right\|_2^2\right\}=\sum_{k=1}^{n} \mathbb{E}_{\mbS}\left\{\left(\sum_{i_{1}=1}^{m_{1}m} \mbS_{j i_{1}}\mbB_{i_{1} i_{2}}\right)^2\right\},
\end{aligned}
\end{equation}
or equivalently, in terms of $\mbG$ in \eqref{St_3},
\begin{equation}
\label{St_120}
\begin{aligned}
\sum_{i_{2}=1}^{n} \mathbb{E}_{\mbG}\left\{\left(\sum_{i_{1}=1}^{k^{\prime}} \mbG^{\top}_{j i_{1}} \mbB_{i_{1}i_{2}}\right)^2\right\}&=\\
\sum_{i_{2}=1}^{n} &\sum_{i_{1}=1}^{k^{\prime}} \mathbb{E}_{\mbG}\left\{\left(\mbG^{\top}_{j i_{1}}\right)^{2}\right\} \mbB^{2}_{i_{1} i_{2}},
\end{aligned}
\end{equation}
with $\mathbb{E}_{\mbG}\left\{\left(\mbG^{\top}_{j i_{1}}\right)^{2}\right\}=1$, which helps to simplify \eqref{St_120} as
\begin{equation}
\label{St_1200}
\begin{aligned}
\sum_{i_{2}=1}^{n} \sum_{i_{1}=1}^{k^{\prime}} \mbB^{2}_{i_{1} i_{2}}=\|\hat{\mbB}\|_{\mathrm{F}}^2,
\end{aligned}
\end{equation}
where $\hat{\mbB}$ is the $k^{\prime}\times n$ submatrix of $\mbB$. Due to the fact that the second term in the right-hand side of \eqref{St_11} is an expectation over the convex function $f(x,y)=x^{2}/y$, we can apply Jensen's inequality as follows:\par\noindent\small
\begin{equation}
\label{St_13}
\begin{aligned}
\mathbb{E}_{\mbS}\left\{ \frac{\left\|\mbS^{\top}\mbB\mbx_i-\mbS^{\top} \mbb \right\|_{\infty}^2}{\left\|\left(\mbS^{\top} \mbB\right)_{j^{\star}_{i}}\right\|_2^2}\right\} \geq \frac{\left(\mathbb{E}_{\mbS}\left\{\left\|\mbS^{\top}\mbB\mbx_i-\mbS^{\top} \mbb \right\|_{\infty}\right\}\right)^2}{\mathbb{E}_{\mbS}\left\{\left\|\left(\mbS^{\top} \mbB\right)_{j^{\star}_{i}}\right\|_2^2\right\}}.
\end{aligned}
\end{equation}
Since $\mbS^{\top}\mbB\mbx_{\star}\preceq \mbS^{\top}\mbb$ and $\mbS^{\top}\mbB\mbx_{i}\succeq\mbS^{\top}\mbb$, one can conclude 
\begin{equation}
\label{NegA}
\left\|\mbS^{\top}\mbB\mbx_i-\mbS^{\top} \mbb \right\|_{\infty}\geq \left\|\mbS^{\top}\mbB\mbx_i-\mbS^{\top}\mbB\mbx_{\star}\right\|_{\infty}.
\end{equation}
It follows from the above that
\begin{equation}
\label{St_130}
\begin{aligned}
\mathbb{E}_{\mbS}\left\{ \frac{\left\|\mbS^{\top}\mbB\mbx_i-\mbS^{\top} \mbb \right\|_{\infty}^2}{\left\|\left(\mbS^{\top} \mbB\right)_{j^{\star}_{i}}\right\|_2^2}\right\} \geq \frac{\left(\mathbb{E}_{\mbS}\left\{\left\|\mbS^{\top} \mbB\left(\mbx_{i}-\mbx_{\star}\right)\right\|_{\infty}\right\}\right)^2}{\|\hat{\mbB}\|_{\mathrm{F}}^2}.
\end{aligned}
\end{equation}
We can additionally take advantage of the estimate for the maximum of independent normal random variables\cite{rebrova2021block},
\begin{equation}
\label{St_14}
\begin{aligned}
\mathbb{E}_{\mbS}\left\{\left\|\mbS^{\top} \mbB\left(\mbx_{i}-\mbx_{\star}\right)\right\|_{\infty}\right\} &=\mathbb{E}_{\mbS}\left\{\max _{t\in[k^{\prime}]}\left\langle \mbs_t, \mbB\left(\mbx_{i}-\mbx_{\star}\right)\right\rangle\right\}\\
&=\mathbb{E}_{\mbG}\left\{\max _{t\in[k^{\prime}]}\left\langle \mbs_t, \hat{\mbB}\left(\mbx_{i}-\mbx_{\star}\right)\right\rangle\right\}\\
&\geq c\|\hat{\mbB}\left(\mbx_{i}-\mbx_{\star}\right)\|_2 \sqrt{\log k^{\prime}},
\end{aligned}
\end{equation}
where $\mbs_{t}$ is the $t$-th column of $\mbS$, $[k^{\prime}]=\left\{1,2,\cdots,k^{\prime}\right\}$, and $c$ is a positive value.
By plugging the inequality \eqref{St_14} into \eqref{St_11}, and using the inequality,
\begin{equation}
\label{St_15}
\left\|\hat{\mbB}\left(\mbx_{i}-\mbx_{\star}\right)\right\|_2^2\geq \sigma^{2}_{\textrm{min}}\left(\hat{\mbB}\right) \left\|\mbx_{i}-\mbx_{\star}\right\|^{2}_{2},
\end{equation}
where $\sigma^{2}_{\textrm{min}}$ is the minimum singular value. Thus, we obtain
\begin{equation}
\label{St_16}
\begin{aligned}
\mathbb{E}\left\{\left\|\mbx_{i+1}-\mbx_{\star}\right\|_2^2\right\}&\leq\left\|\mbx_i-\mbx_{\star}\right\|_2^2-\frac{c\|\mbB\left(\mbx_{\star}-\mbx_i\right)\|_2^2\log k^{\prime}}{\|\mbB\|_{\mathrm{F}}^2}\\\quad&\leq\left\|\mbx_i-\mbx_{\star}\right\|_2^2-\frac{c \sigma_{\min }^2(\hat{\mbB}) \log k^{\prime}}{\|\hat{\mbB}\|_{\mathrm{F}}^2}\left\|\mbx_i-\mbx_{\star}\right\|_2^2\\\quad &\leq \left(1-\frac{c \sigma_{\min }^2(\hat{\mbB}) \log k^{\prime}}{\|\hat{\mbB}\|_{\mathrm{F}}^2}\right)\left\|\mbx_i-\mbx_{\star}\right\|_2^2,
\end{aligned}
\end{equation}
which can be recast as the following \emph{convergence rate}, after $K$ updates:
\begin{equation}
\label{St_17}
\begin{aligned}
\mathbb{E}\left\{\left\|\mbx_{i+1}-\mbx_{\star}\right\|_2^2\right\}\leq \left(1-\frac{c \sigma_{\min }^2(\hat{\mbB}) \log k^{\prime}}{\|\hat{\mbB}\|_{\mathrm{F}}^2}\right)^{K}\left\|\mbx_0-\mbx_{\star}\right\|_2^2.
\end{aligned}
\end{equation}
%--------------------------------------------------------------------------------
\begin{figure}[t]
	\centering
	\subfloat[]
		{\includegraphics[width=0.45\columnwidth]{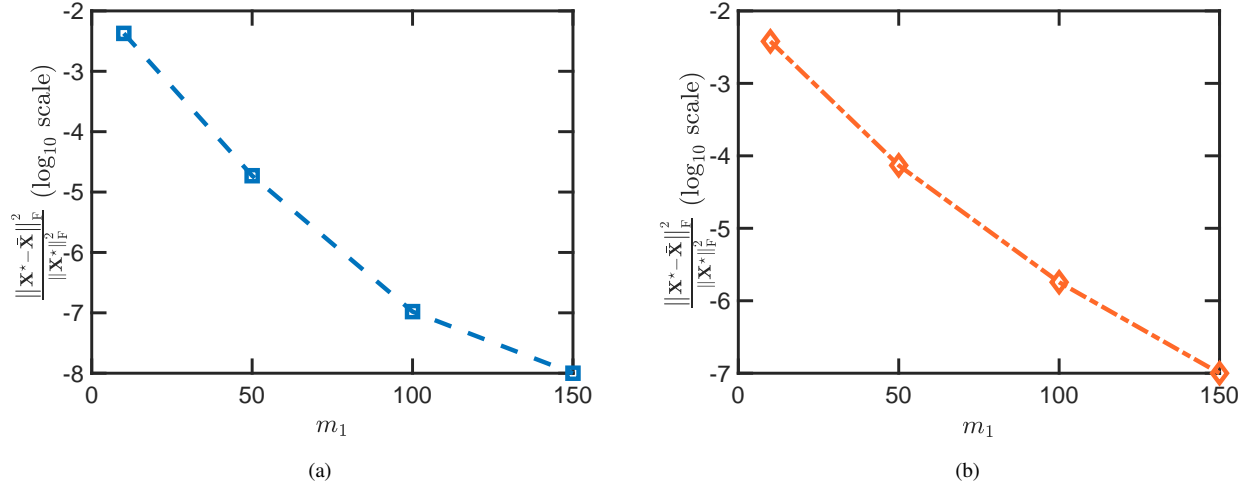}}\quad
	\subfloat[]
		{\includegraphics[width=0.45\columnwidth]{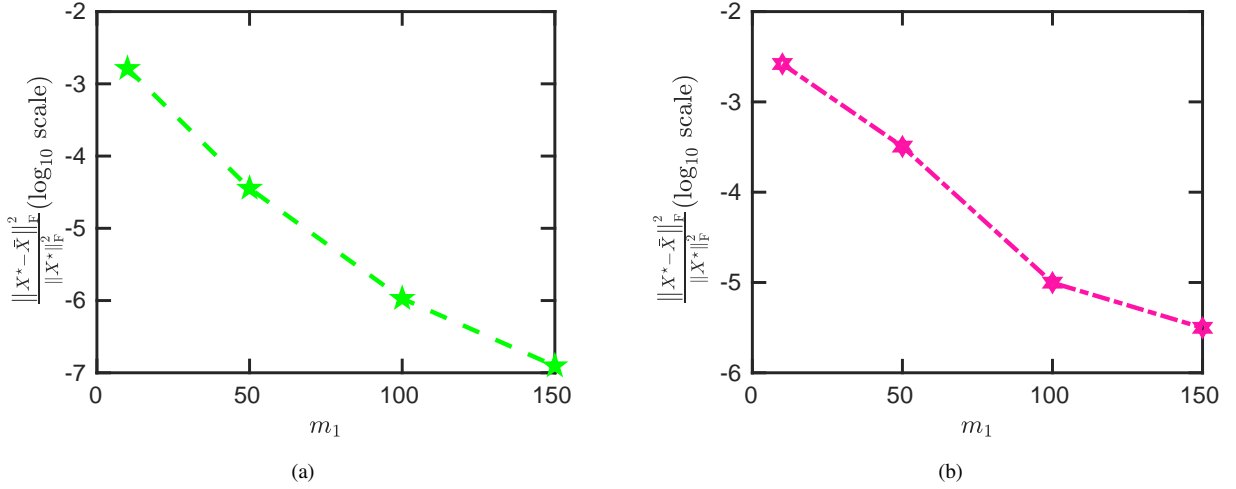}}
	\caption{Average NMSE for the error between the desired  $\mbX^{\star}$ and its recovered version $\bar{\mbX}$ for various numbers of time-varying sampling threshold sequences $m_{1}\in\left\{10,50,100,150\right\}$ when the block SKM is utilized in with (a) $\|\mathbf{x}\|_{0}=5$, (b) $\|\mathbf{x}\|_{0}=10$. %The results are obtained for $\mbx\in\mathbb{R}^{64}$ and rank one matrices $\left\{\mbA_{j}^{64\times 64}\right\}_{j=1}^{5000}$.
 }
	\label{figure_1}
\end{figure}
%--------------------------------------------------------------------------------
%--------------------------------------------------------------------------------
\begin{figure}[t]
	\centering
	\subfloat[]
		{\includegraphics[width=0.45\columnwidth]{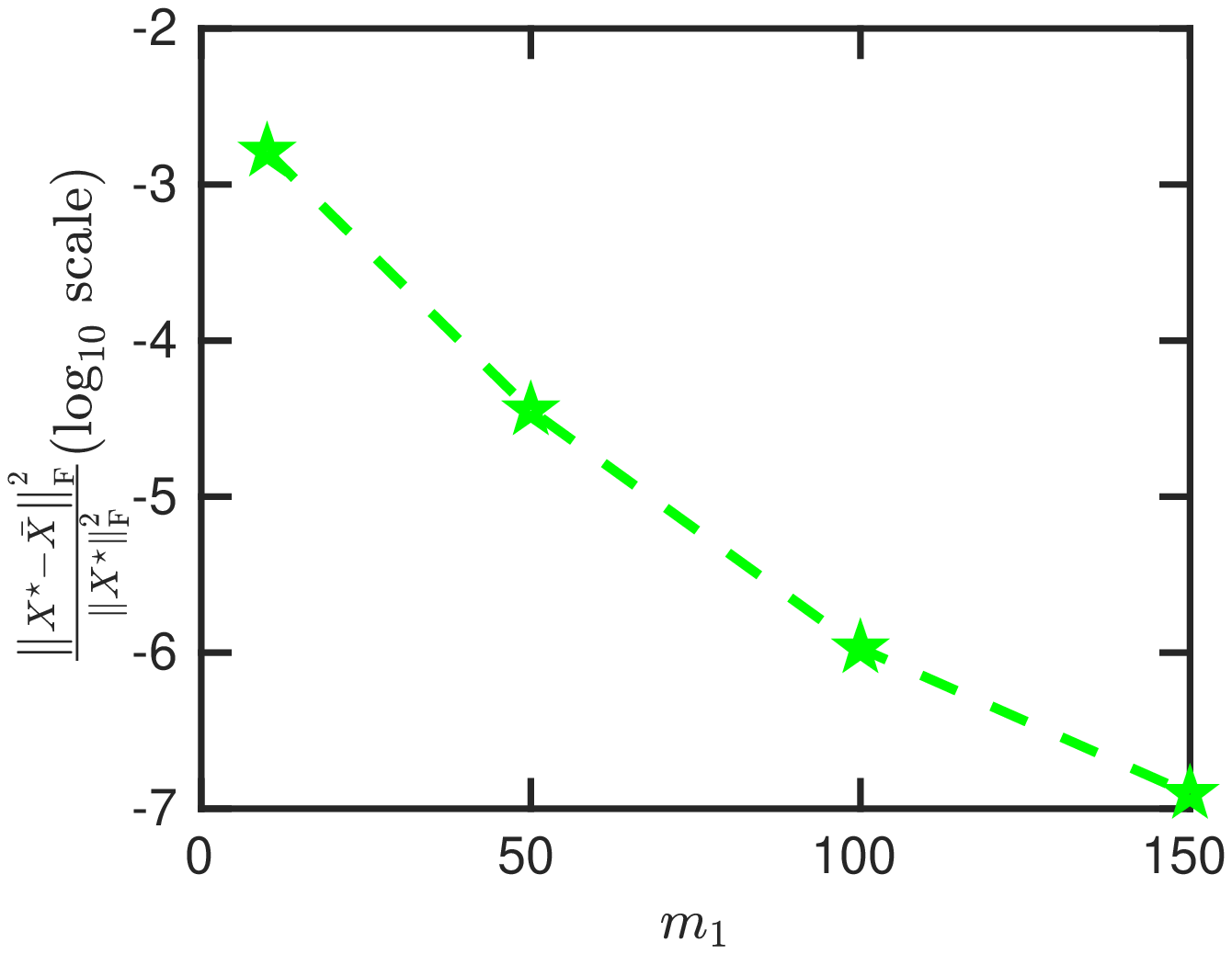}}\quad
	\subfloat[]
		{\includegraphics[width=0.45\columnwidth]{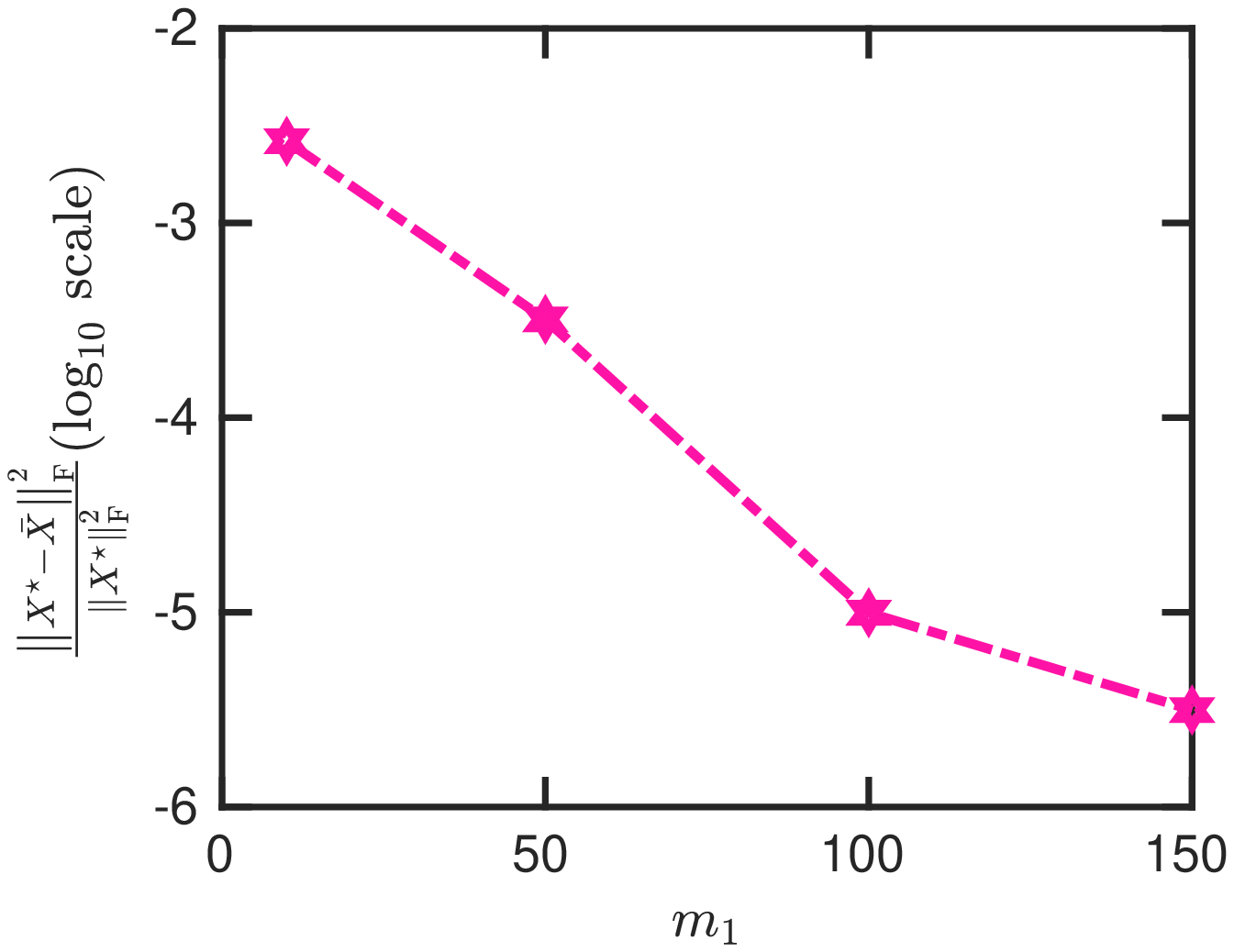}}
	\caption{Average NMSE for the error between the desired  $\mbX^{\star}$ and its recovered version $\bar{\mbX}$ for various numbers of time-varying sampling threshold sequences $m_{1}\in\left\{10,50,100,150\right\}$ when the block SKM is utilized in with (a) $\|\mathbf{x}\|_{0}=5$, (b) $\|\mathbf{x}\|_{0}=10$. %The results are obtained for $\mbx\in\mathbb{R}^{64}$ and full rank matrices $\left\{\mbA_{j}^{64\times 64}\right\}_{j=1}^{5000}$.
 }
	\label{figure_2}
\end{figure}
%--------------------------------------------------------------------------------

\section{Numerical Results}
\label{num}
In this section, at first, we numerically scrutinize the capability of the block SKM in the one-bit QCS problem by evaluating the squared Frobenius norm of the error between the desired matrix $\mbX^{\star}$ and its estimate $\bar{\mbX}$, normalized by the squared Frobenius norm of the desired matrix:
\begin{equation}
\label{eq:4000}
\mathrm{NMSE}\triangleq\frac{\left\|\mbX^{\star}-\bar{\mbX}\right\|^{2}_{\mathrm{F}}}{\left\|\mbX^{\star}\right\|^{2}_{\mathrm{F}}}.
\end{equation}
The input signal $\mathbf{x}\in\mathbb{R}^{64}$, is considered to be a sparse signal with (i) $\|\mathbf{x}\|_{0}=5$, and (ii) $\|\mathbf{x}\|_{0}=10$. To choose the time-varying sampling thresholds, we consider the framework presented in \cite{laska2011trust}, which relies on knowledge of the dynamic range of the measurements $\mby$. Assume $\beta_{\mby}=\left\|\mby\right\|_{\infty}$ denotes the dynamic range of the measurements. Then, herein we generate the time-varying sampling thresholds as $\left\{\mathbf{\uptau}^{(\ell)}\sim \mathcal{N}\left(\mathbf{0},\frac{\beta_{\mby}^{2}}{9}\mathbf{I}_{5000}\right)\right\}_{\ell=1}^{m_{1}}$. Each sensing matrix is generated based on $\mbA_{j}=\mba_{j}\mba^{\mathrm{H}}_{j}$, where $\mba_{j}\sim \mathcal{N}\left(\mathbf{0},\mbI_{64}\right)$. 
We solve the overdetermined one-bit QCS polyhedron in (\ref{eq:8n}) via the Block SKM for the number of time-varying sampling threshold sequences $m_{1}\in \left\{10, 50, 100, 150\right\}$. Fig.~\ref{figure_1} appears to confirm the possibility of recovering the desired matrix $\mbX^{\star}$ in the one-bit QCS polyhedron (\ref{eq:8n}) by applying Block SKM. As expected, the performance of the recovery will be significantly enhanced as the number of time-varying sampling threshold sequences grows large. The reason behind this observation is the sample abundance condition which has been initially analyzed and proved in \cite[Theorem 1]{eamaz2022one} and extended to another sampling scheme in \cite{eamaz2022uno}. Note that the results in Fig.~\ref{figure_1} are averaged over $15$ experiments. To examine the performance of the proposed algorithm for the full rank $\mbA_{j}$ scenario, we generate a full rank $\mbA_{j}\in\mathbb{R}^{64\times 64}$ where its entries are i.i.d normal random variables. Similarly, we generate time-varying sampling thresholds as $\left\{\mathbf{\uptau}^{(\ell)}\sim \mathcal{N}\left(\mathbf{0},\frac{\beta_{\mby}^{2}}{9}\mathbf{I}_{5000}\right)\right\}_{\ell=1}^{m_{1}}$. Fig.~\ref{figure_2} illustrates the recovery performance of the Block SKM in this case while preserving the property of boosting the recovery error as the number of time-varying sampling thresholds grows large. Each data point in Fig.~\ref{figure_2} is averaged over $15$ experiments. Note that the algorithm proposed employs only low-resolution (one-bit) samples, but capitalizes on their abundance to converge to the global solution with heightened precision as the quantity of one-bit samples increases.
%--------------------------------------------------------------------------------
\begin{figure}[t]
	\center{\includegraphics[width=0.45\textwidth]{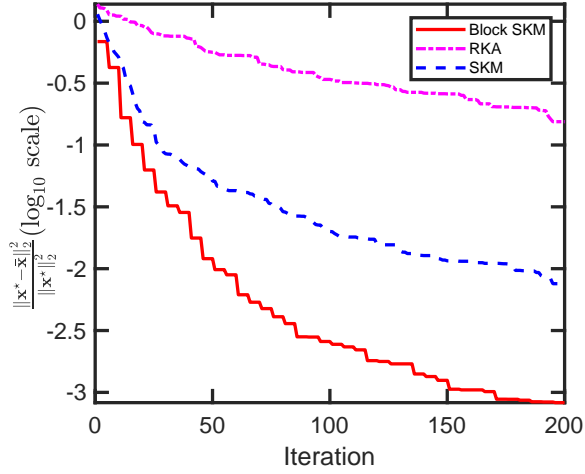}}
	\caption{Comparing the recovery performance of the proposed Kaczmarz-based algorithm, namely the Block SKM, with that of SKM and RKA in terms of NMSE for a linear system of inequalities. %which is created by applying one-bit sampling to a system of linear equalities $\mbB\mbx=\mby$, where $\mbB\in\mathbb{R}^{100\times 10}$, $\mathbf{x}\in\mathbb{R}^{10}$, $\mathbf{y}\in\mathbb{R}^{100}$, and the number of time-varying sampling thresholds $m_{1}=40$.
 }
	\label{figure_3}
\end{figure}
%--------------------------------------------------------------------------------
\begin{table} [t]
\caption{Comparing CPU times and $\operatorname{NMSE}$ of Block SKM and GESPAR.}
\centering
\begin{tabular}{ | c | c | c | c |}
\hline
\text {Algorithm} & \text {$m^{\star}$} & \text {CPU time (s)} & \text {$\operatorname{NMSE}$} \\[0.5 ex]
\hline \hline
\text{Block SKM (one-bit)} & $5000$ &  $0.0026$ & $3.1072e-7$ \\[1 ex]
\hline
\text{GESPAR} & $128$ &  $0.0041$ & $2.4382e-5$ \\[1 ex]
\hline
\end{tabular}
\label{table_1}
\end{table}
Moreover, we numerically compare the RKA\cite{leventhal2010randomized}, SKM\cite{de2017sampling}, and our proposed Block SKM in linear systems of inequalities. We apply one-bit sampling to a system of linear equalities $\mbB\mathbf{x}=\mathbf{y}$, resulting in the creation of its corresponding system of linear inequalities as described in (\ref{eq:8}). Herein, we consider 
%the number of time-varying sampling threshold sequences is $m_{1}=40$, 
$\mbB\in\mathbb{R}^{100\times 10}$, $\mathbf{x}\in\mathbb{R}^{10}$, and $\mathbf{y}\in\mathbb{R}^{100}$. Each row of $\mbB$ is generated as $\mathbf{b}_{j}\sim\mathcal{N}\left(\mathbf{0},\mbI_{10}\right)$. Also, the desired signal $\mathbf{x}$ is generated as $\mathbf{x}\sim\mathcal{N}\left(\mathbf{0},\mbI_{10}\right)$. Accordingly, we generate time-varying sampling thresholds as $\left\{\mathbf{\uptau}^{(\ell)}\sim \mathcal{N}\left(\mathbf{0},\frac{\beta_{\mby}^{2}}{9}\mathbf{I}_{100}\right)\right\}_{\ell=1}^{m_{1}}$ for $m_{1}=40$. The performance of the RKA, SKM, and Block SKM is illustrated in Fig.~\ref{figure_3}. The results show that the Block SKM outperforms the other two approaches, delivering a faster recovery and higher accuracy in the recovery of the desired signal $\mathbf{x}$.
%It can be seen that the Block SKM leads to a faster recovery and better accuracy in the recovery of the desired signal $\mathbf{x}$ compared to the other two approaches. 
The normalized mean square error for the signal is defined as $\operatorname{NMSE}\triangleq\frac{\left\|\mathbf{x}_{\star}-\bar{\mathbf{x}}\right\|_{2}^{2}}{\left\|\mathbf{x}_{\star}\right\|_{2}^{2}}$, where $\mathbf{x}_{\star}$ and $\bar{\mathbf{x}}$ denote the true discretized signal and its recovered version, respectively. The NMSE results in Fig.~\ref{figure_3} are averaged over $15$ experiments.

To further investigate the efficacy of the proposed algorithm in QCS, we compare our proposed approach with the well-known GESPAR approach with the initialization algorithm proposed in \cite{shechtman2014gespar} in terms of NMSE and CPU time. As presented in Table~\ref{table_1}, Block SKM outperforms GESPAR in terms of both NMSE and CPU time. The results are obtained for $\mathbf{x}\in\mathbb{R}^{64}$ when the optimal number of samples are utilized, and where $m^{\star}=2n=128$ (high-resolution samples) and $m^{\star}=5000$ (one-bit samples) are considered for the high-resolution method and one-bit QCS, respectively. Herein, the optimality of sample sizes means that the number of samples utilized by algorithms leads to their best performance (up to global phase), i.e. satisfying the criterion $\left\|\mbx_{\star}-\bar{\mbx}\right\|_{2}^{2}\leq 5\times 10^{-5}\left\|\mbx_{\star}\right\|_{2}^{2}$. By this comparison, we remove the burden of the large number of samples from the GESPAR to fairly compare their optimal shape deploying incomplete measurements with that of the Block SKM. Note that the signal of interest is obtained from $\bar{\mbX}=\bar{\mbx}\bar{\mbx}^{\mathrm{H}}$, where the signal is the largest eigenvector of the recovered matrix.

\section{Conclusion}
\label{conclude}
We propose taking advantage of the abundant number of samples available in one-bit sampling with time-varying thresholds to efficiently and globally solve the quadratic compressed sensing problem. In particular, a state-of-the-art randomized Kaczmarz algorithms is proposed to find the desired signal inside the emerging confined feasible regions, named the one-bit polyhedron, with an enhanced convergence rate. The numerical results showcased the effectiveness of the proposed approaches for the quadratic compressed sensing problem.

%\newpage
\bibliographystyle{IEEEtran}
\bibliography{refs}

\end{document}